%% file: _main.tex
\documentclass[journal]{vgtc}              %

\usepackage[table]{xcolor}

\input{preamble}

\title{AnnoBench: A Benchmark for Visualization Annotation Generation}

\usepackage{graphicx} %

\newcommand{\cofirstglyph}{%
  \scalebox{1.3}{\mbox{\normalfont\bfseries *}}%
}

\newcommand{\cofirstmark}{%
  \texorpdfstring{\raisebox{0.75ex}{\cofirstglyph}}{}%
}

\author{%
  \authororcid{Md Rahat-uz-Zaman}{0000-0001-6728-7569}\cofirstmark,
  \authororcid{Md Dilshadur Rahman}{0009-0008-5467-615X}\cofirstmark,
  \authororcid{Andrew McNutt}{0000-0001-8255-4258},
  and \authororcid{Paul Rosen}{0000-0002-0873-9518}
  \texorpdfstring{%
    \\[-0.05em]
    {\small \cofirstglyph\;equal contribution}%
  }{}%
}

\authorfooter{
  \item Md Rahat-uz-Zaman is with the University of Utah.
    E-mail: rahatzamancse@gmail.com.

  \item Md Dilshadur Rahman is with the University of Utah.
    E-mail: dilshadur@sci.utah.edu.

  \item Andrew McNutt is with the University of Utah.
    E-mail: andrew.mcnutt@utah.edu.

  \item Paul Rosen is with the University of Utah.
    E-mail: paul.rosen@utah.edu.
}

\definecolor{colFair}{HTML}{fa9c5a}
\definecolor{colModerate}{HTML}{f5d27d}
\definecolor{colSubstantial}{HTML}{b0d670}
\definecolor{colPerfect}{HTML}{0a6e37}

\abstract{
\input{00-abstract}

}

\keywords{Annotations, Visualizations, Benchmark study, Large Language Model}

\input{00-teaser}

\renewcommand{\manuscriptnotetxt}{}

\graphicspath{{figs/}{figures/}{pictures/}{images/}{./}} %

\usepackage{tabularx}
\usepackage{booktabs}
\usepackage{multirow}
\usepackage{setspace}

\usepackage{array}
\usepackage{ragged2e}
\usepackage{caption}
\usepackage{tikz}

\usepackage{mathptmx}                  %

\begin{document}

\setstretch{0.975}

\maketitle
\input{01-introduction}

\input{02-background}

\input{03-design-principle}

\input{04-eval-framework}

\input{05-annobench}

\input{06-annobench-system}

\input{07-benchmark-evaluation}

\input{08-discussion}

\input{09-conclusion}

\acknowledgments{%
  We used the generative AI tool GPT--5 by OpenAI to edit, paraphrase, and restructure the text, which was later reviewed and revised. This work was supported in part by a grant from NSF (\#2402719).%
}

\bibliographystyle{abbrv-doi-hyperref}

\bibliography{_bib_clean}

\newpage
\appendix %

\input{10-appendix}

\end{document}

%% file: preamble.tex
\newcommand{\systemLink}{\ourhref{https://tinyurl.com/annobench}{tinyurl.com/annobench}}
\newcommand{\sourceLink}{\ourhref{https://anonymous.4open.science/r/AnnoBench}{anonymous.4open.science/r/AnnoBench}}

\definecolor{chart-description-0}{HTML}{EDEDED}
\definecolor{chart-description-1}{HTML}{CBCBCB}
\definecolor{chart-description-2}{HTML}{FFEBD8}
\definecolor{chart-description-3}{HTML}{CAEFCA}
\definecolor{chart-description-4}{HTML}{FFFFD4}
\definecolor{chart-description-5}{HTML}{95C1EA}

\definecolor{precisecolor}{RGB}{52,152,219}
\definecolor{relevantcolor}{RGB}{46,204,113}
\definecolor{consistentcolor}{RGB}{155,89,182}
\definecolor{visiblecolor}{RGB}{241,196,15}
\definecolor{preservecolor}{RGB}{230,126,34}

\newcommand{\etal}{et al.}

\newcommand{\eg}{{e.g.,}}

\newcommand{\figref}[1]{\hyperref[#1]{Fig.~\ref*{#1}}}
\newcommand{\secref}[1]{\hyperref[#1]{Sec.~\ref*{#1}}}

\newcommand{\paraheadd}[1]
{%
    \vspace{0.07in}%
    \noindent%
    \textbf{\textit{#1}}%
}

\newcommand{\parahead}[1]{\paraheadd{#1.}}

\usepackage[normalem]{ulem}

\renewcommand{\paragraph}[1]{\textbf{#1:}}

\usepackage{xspace}
\usepackage[pagebackref,bookmarks]{hyperref}

\definecolor{steelblue}{RGB}{70, 130, 180}

\hypersetup{
    colorlinks=true, 
    linkcolor=steelblue, 
    urlcolor=steelblue,
    citecolor=steelblue
}

\let\oldhref\href

\newcommand{\ourhref}[2]{\oldhref{#1}{\textcolor{steelblue}{#2~$\nearrow$}}}

\definecolor{annoPrecise}{HTML}{C4CDE1}
\definecolor{annoRelevant}{HTML}{E0D7C4}
\definecolor{annoVisible}{HTML}{E0C8C4}
\definecolor{annoPreservant}{HTML}{A3C8BF}
\definecolor{annoConsistant}{HTML}{AAD1B1}

\newcommand{\preciseTag}{%
  \begingroup
  \setlength{\fboxsep}{1.5pt}%
  \colorbox{annoPrecise}{\textbf{Precise}}%
  \endgroup
}

\newcommand{\relevantTag}{%
  \begingroup
  \setlength{\fboxsep}{1.5pt}%
  \colorbox{annoRelevant}{\textbf{Relevant}}%
  \endgroup
}

\newcommand{\visibleTag}{%
  \begingroup
  \setlength{\fboxsep}{1.5pt}%
  \colorbox{annoVisible}{\textbf{Visible}}%
  \endgroup
}

\newcommand{\preservantTag}{%
  \begingroup
  \setlength{\fboxsep}{1.5pt}%
  \colorbox{annoPreservant}{\textbf{Preservant}}%
  \endgroup
}

\newcommand{\consistantTag}{%
  \begingroup
  \setlength{\fboxsep}{1.5pt}%
  \colorbox{annoConsistant}{\textbf{Consistent}}%
  \endgroup
}

\newcommand{\precise}{precise}
\newcommand{\visible}{visible}
\newcommand{\relevant}{relevant}
\newcommand{\preservant}{preservant}
\newcommand{\consistent}{consistent}

%% file: 00-abstract.tex
Annotation is among the most demanding visualization tasks to automate, as it simultaneously requires correctly navigating visual, semantic, and stylistic constraints. Failure to meet any of these conditions severely undermines the utility of an annotation, rendering it challenging to read, inaccurate, or visually discordant. Despite a growing body of annotation tools and automations, no existing benchmark or evaluation framework tests whether these conditions are met because of their scope and annotation not being the focus of their studies.
We introduce AnnoBench, a benchmark for visualization annotation that materializes the inherent challenges of this domain in a structured and testable manner. AnnoBench pairs visualizations from professional data journalism and visualization galleries with annotation tasks, spanning four representation formats, five chart description conditions, and two prompt specification levels.
The benchmark is executed via VLM-as-a-judge, using models aligned with manual human assessment.
We evaluate the benchmark via four one-factor-at-a-time experiments, exploring the effects of input representation, semantic context, and prompt specificity, and model selection on annotation quality.
This work provides a foundation for advancing annotation automation, tooling, and visualization-generation pipelines.

%% file: 00-teaser.tex
\teaser{
  \centering
  \includegraphics[width=\linewidth, alt={A screenshot of AnnoBench Browser system and demonstration of AnnoBench dataset dimensions, and annotation evaluation with experimental results. }]{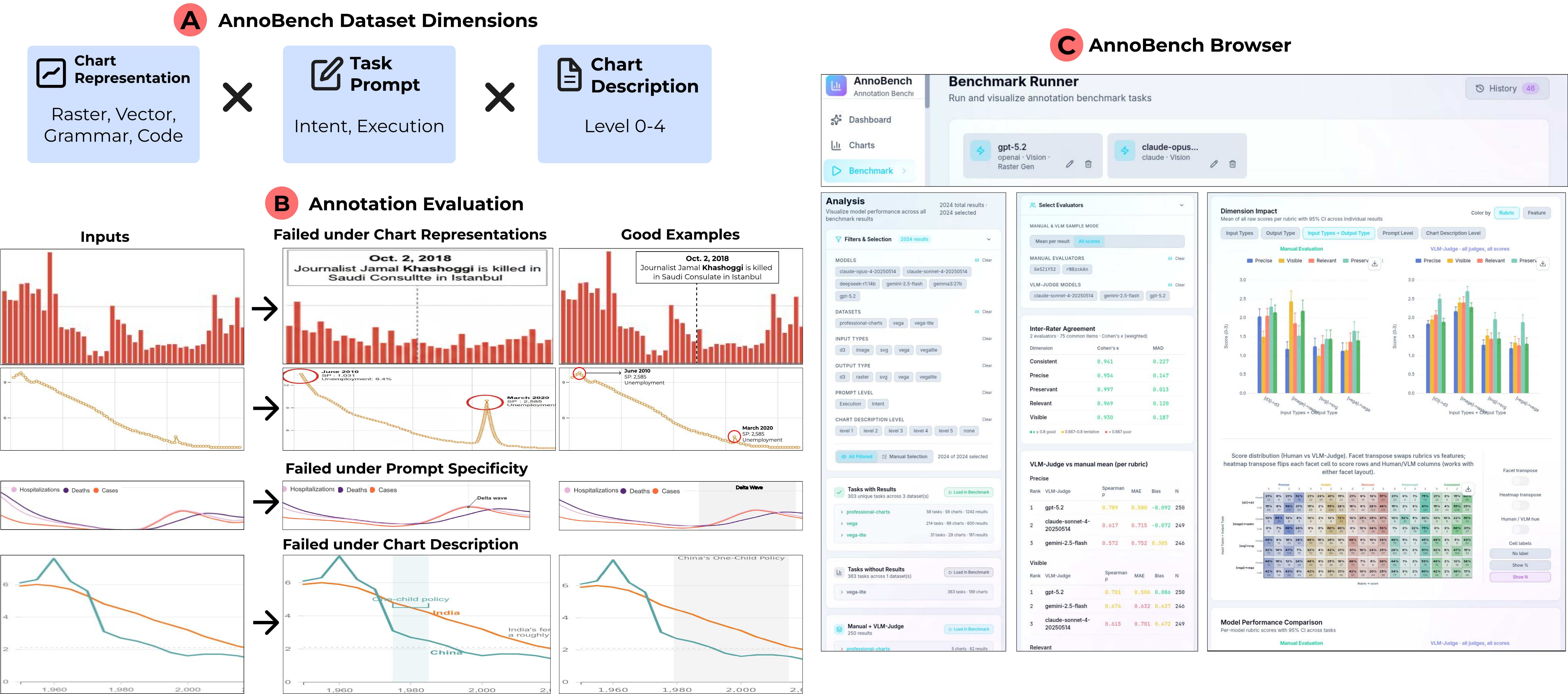}
  \caption{%
We introduce a three-dimensional benchmark for evaluating chart annotation in large language models (A), along with a five-dimensional evaluation framework to assess annotation quality. 
Using AnnoBench, we reveal systematic failure modes in modern models, including incorrect targeting, data distortion, and avoidance of certain annotation types (middle column in (B)). To support reproducible experimentation and comparison, we provide an interactive web-based interface for configuring, running, and analyzing benchmark trials (C).
  }
  \label{fig:teaser}
}

%% file: 01-introduction.tex
\section{Introduction}
\label{sec:intro}
Annotations are among the most critical components of a visualization~\cite{rahman2025survey}, guiding readers through content via careful placement and contextualization that must be both graphically and semantically appropriate. 
For these reasons, annotation is among the most demanding visualization tasks to automate. An annotation must target the correct visual element~\cite{kong2012graphical, ren2017chartaccent}, assert only what the underlying encoding supports~\cite{rahman2025survey, hullman2013contextifier}, and fit within the visualization’s existing layout without occluding evidence~\cite{kittivorawong2020fastlabels}. These conditions compound in ways that make failures hard to anticipate. For example, an annotation labeling a bar as ``peak revenue'' is incorrect if that bar encodes the second-highest value, even if the annotation text is well-phrased and clearly positioned. Conversely, a factually correct annotation may have no viable placement in a densely labeled region without occluding the data it references.  
Annotations are visually anchored to the elements they describe, and so such errors can be difficult to detect~\cite{lisnic2023misleading, Arlene2022annotating} and can shift reader conclusions in otherwise accurate visualizations~\cite{stokes2022striking}.

A range of tools have taken up this challenge, including interactive environments~\cite{ren2017chartaccent}, semi-automatic insight externalization systems~\cite{chen2010click2annotate}, automated generation pipelines~\cite{hullman2013contextifier, gao2014newsviews}, and specification grammars~\cite{rahman2025annogram, chen2025chartmark}. Prior work has also characterized annotation forms and targets~\cite{ren2017chartaccent, rahman2024qualitative, kong2012graphical}, the communicative purposes annotations serve~\cite{rahman2024qualitative, hullman2013contextifier, rahman2024exploring}, and their measurable effects on reader comprehension and bias perception~\cite{stokes2023role, stokes2022striking, ottley2019curious, fan2024understanding, Arlene2022annotating}. Yet across both the tools and the research literature, evaluable conditions for annotation correctness remain undefined~\cite{rahman2025survey, spivak2025fiction}. Without such conditions, annotation quality can only be judged informally, and failures cannot be systematically identified or compared across systems. Existing benchmarks for chart-related tasks evaluate chart comprehension~\cite{lee2016vlat, bendeck2024empirical, hong2025vlat}, text answers~\cite{masry2022chartqa, methani2020plotqa}, natural-language descriptions~\cite{tang2023vistext, kantharaj2022charttotext}, or chart specifications~\cite{chen2024viseval, rahman2025text2vis}, but none considers annotation quality.
These problems further compound as we increasingly draw on generative AI models for visualization construction support~\cite{li2024we, wang2025dataformulator2}. 

We address these gaps with an evaluation framework, discussed in \autoref{sec:design}, that operationalizes three necessary conditions for annotation correctness---target correctness, task accuracy, and spatial feasibility---alongside rubric dimensions for stylistic coherence and chart preservation. We instantiate this framework as \textbf{AnnoBench}, discussed in~\autoref{sec:dataset} and \autoref{sec:annobench_browser}, a benchmark that evaluates Large Language Models (LLMs) and Vision Language Models (VLMs) under controlled variation in chart representation, semantic description, task specification, and model class. This design enables us to analyze how access to chart structure, contextual support, and prompt specificity affects annotation quality across different model types. 
We evaluate AnnoBench through a set of controlled experiments, reported in~\autoref{sec:evaluation}. 

\vspace{3pt}
\noindent
We make three contributions:
\begin{itemize}[leftmargin=0.1in]

\item \textbf{Evaluation framework}: 
An evaluation framework for chart annotation that distinguishes per-instance correctness criteria from broader rubric dimensions. We operationalize this framework as a five-dimensional rubric consisting of target correctness, task accuracy, spatial feasibility, stylistic coherence, and chart preservation, enabling comparable evaluation of annotations produced by visualization tools, grammars, and generation pipelines.

\item \textbf{AnnoBench dataset}: 
A benchmark dataset of about 342 charts drawn from four source collections, with over 650 annotation tasks defined at both intent and execution levels. The dataset spans six formats across four representation families (D3, Vega, Vega-Lite, ggplot2, SVG, and raster) and five chart description conditions: no description, construction and encoding, statistical and relational, perceptual and cognitive, and contextual and domain-specific. This design enables controlled analysis of how representation, semantic context, and task specification affect annotation performance. 

\item \textbf{Benchmark system}: 
An end-to-end benchmarking pipeline for running annotation evaluations across built-in and custom models, with format-aware rendering, automated VLM-as-a-judge scoring, and human evaluation support. This system provides reusable infrastructure for applying the benchmark to future annotation methods under a consistent evaluation setup.

\end{itemize}

With this work, we establish chart annotation as a more explicit and testable target for evaluation, enabling more rigorous analysis of automated systems, informing the development of future annotation methods, tools, and evaluation practice, and ultimately helping improve the quality of annotations more generally. Our source code is available at \sourceLink{}, and an interactive browser for the benchmark is available at \systemLink{}.

%% file: 02-background.tex
\section{Related Work}
\label{sec:related-work}

We situate AnnoBench at the intersection of two bodies of work: 
visualization research on the structure and effects of annotations, 
and benchmark evaluation of LLMs and VLMs on visualization tasks.

\subsection{Annotations in Visualization}
\label{sec:related-work:annotations}

Research on annotations addresses three aspects 
relevant to AnnoBench: how annotations are structured, how they 
are formalized, and how their design affects readers.

\parahead{Annotation structure and formalization}
Annotations are chart-bound additions that attach text or graphical 
cues to specific chart elements to explain, emphasize, compare, or 
contextualize data features~\cite{kong2012graphical, 
rahman2024qualitative, rahman2023exploring}. Two structural dimensions 
organize this space---annotation forms and annotation targets. Form refers to the graphical vehicle of the 
annotation, such as text labels, connectors, enclosures, and 
highlights. Target refers to the chart element the annotation 
addresses, such as individual data items, regions, or broader chart 
structures~\cite{kong2012graphical, ren2017chartaccent}. Several 
systems and grammars make these dimensions explicit. For instance, ChartAccent 
organizes annotation design around paired form and target 
choices~\cite{ren2017chartaccent}. Contextifier distinguishes 
annotations that restate visible patterns from those that introduce 
external context and models anchoring at multiple levels of the 
visualization~\cite{hullman2013contextifier}. Recent grammar-based 
work treats annotation semantics, targets, and placement as 
first-class components of visualization 
specifications~\cite{rahman2025annogram, chen2025chartmark}. In 
practice, a single form is often insufficient to unambiguously 
identify its target. Professional graphics, therefore, frequently 
combine forms into ensembles such as connector+text and 
enclosure+connector+text, with cue selection varying by task and 
display context~\cite{rahman2024qualitative, rahman2023exploring, 
kong2017internal}. Observed practice suggests that form-target choices remain dependent on task, display density, and communicative purpose~\cite{rahman2024qualitative, rahman2023exploring, kong2017internal}.

\parahead{Effects on readers}
Annotation design shapes how information is attached to a chart and, 
by extension, what readers notice, how they navigate the display, and 
what they take away. Whether an annotation restates a pattern visible 
in the chart or introduces external context constrains both the 
appropriate form and the claims it can legitimately 
make~\cite{hullman2013contextifier, rahman2024qualitative}. 
Presentation choices more broadly influence memorability, 
comprehension, and 
recall~\cite{borkin2013makes, borkin2015beyond, bateman2010useful}, 
and within annotated views, specifically, placement, linking, and layout affect gaze behavior, comprehension of referenced evidence, 
and cognitive load~\cite{bryan2020analyzing, ottley2019curious, 
zhi2019linking, bancilhon2023combining}. These effects are 
design-dependent. Augmentation choices vary by form, and cue 
effectiveness changes across presentation 
settings~\cite{chun2020giving, guo2024we, kong2019understanding}. 
The same mechanisms can also distort interpretation. Text amount, framing, and wording shift reader takeaways and perceived bias even when the underlying chart is accurate~\cite{stokes2022striking, stokes2023role, fan2024understanding, Arlene2022annotating, lisnic2023misleading}. Because these effects 
are both design-dependent and consequential, annotation errors are 
not merely stylistic. Incorrect placement, misleading phrasing, or semantically inappropriate context can change what readers conclude 
from an otherwise accurate chart.

Prior work provides a vocabulary for describing annotations,
formal tools for specifying them, and evidence that annotation
design shapes reader outcomes. But it does not establish whether
a particular annotation is correct for a given chart and task.
Doing so requires verifiable per-instance conditions: whether an
annotation targets the intended visual element, asserts only what
the encoding supports, and can be placed without occluding
evidence. Design guidelines, reader outcome measures, and
editorial conventions operate at a broader level and therefore do
not provide such conditions. Rahman et al.~\cite{rahman2025survey}
and Spivak \& Tory~\cite{spivak2025fiction} identify this absence
as an open problem. In \autoref{sec:design}, we address it by
defining per-instance correctness criteria and operationalizing
them in a structured evaluation framework.

\subsection{LLM and VLM Benchmarks for Visualization}
\label{sec:related-work:llms}

A growing body of benchmarks evaluates LLMs and VLMs on chart-related 
tasks, spanning chart comprehension, natural language description, and chart generation and editing.

\parahead{Chart comprehension benchmarks}
Chart comprehension benchmarks take the form of question answering:
a model receives a chart and a natural language question, then is
evaluated on whether it can produce a short, correct answer. Chart
QA benchmarks cover numerical, relational, encoding-level, and
multi-chart reasoning~\cite{methani2020plotqa, masry2022chartqa,
kafle2018dvqa, mukherjee2025encqa, zhu2025multichartqa}.
Visualization literacy instruments assess chart reading through
controlled item sets with known answers~\cite{lee2016vlat,
pandey2023mini, ge2023calvi}, and recent work uses them to probe
VLM generalization under systematic chart
modifications~\cite{hong2025vlat, pandey2025benchmarking}. These
benchmarks surface one capability central to chart annotation: precise
identification and reading of chart elements. Evaluations report
systematic difficulty in locating and reading particular values or
marks~\cite{bendeck2024empirical, mukherjee2025encqa}, a
prerequisite for annotation placement, even though these benchmarks do not evaluate chart modification itself.

\parahead{Chart captioning and generation benchmarks}
Captioning benchmarks evaluate chart-level natural language
descriptions. For instance, VisText represents each chart as a
rasterized image, a backing data table, and a scene graph, and studies
how representation type and semantic level affect captioning
quality~\cite{tang2023vistext}. Chart-to-text benchmarks address chart
summarization, and Lundgard and Satyanarayan~\cite{lundgard2022accessible}
show that what counts as an adequate description depends on the reader
and the task. NL2VIS benchmarks address chart construction from
natural language~\cite{luo2021nvbench, chen2024viseval,
rahman2025text2vis}. Chart editing benchmarks instead require
modifying an existing chart under textual or multimodal
instructions~\cite{zhao2025chartedit, yang2025chartm3,
zou2024vgbench}. Across these benchmarks, the output is a description,
a specification, or a modified chart. Annotation introduces a more
specific evaluation target: a chart-bound addition that must select the
intended referent, state a supported claim, and fit within the chart
layout.

Prior benchmark work suggests three dimensions that are especially relevant to annotation evaluation: chart representation, semantic context, and task specification. Annotation depends on all three simultaneously because it requires a system to identify the intended referent, determine what the chart supports, and incorporate the requested addition into the visualization. \autoref{sec:design} develops these dimensions as the basis for the benchmark design.

%% file: 03-design-principle.tex
\section{Benchmark Design and Evaluation Framework}
\label{sec:design}

\begin{figure*}[ht]
    \centering
    \includegraphics[width=\linewidth]{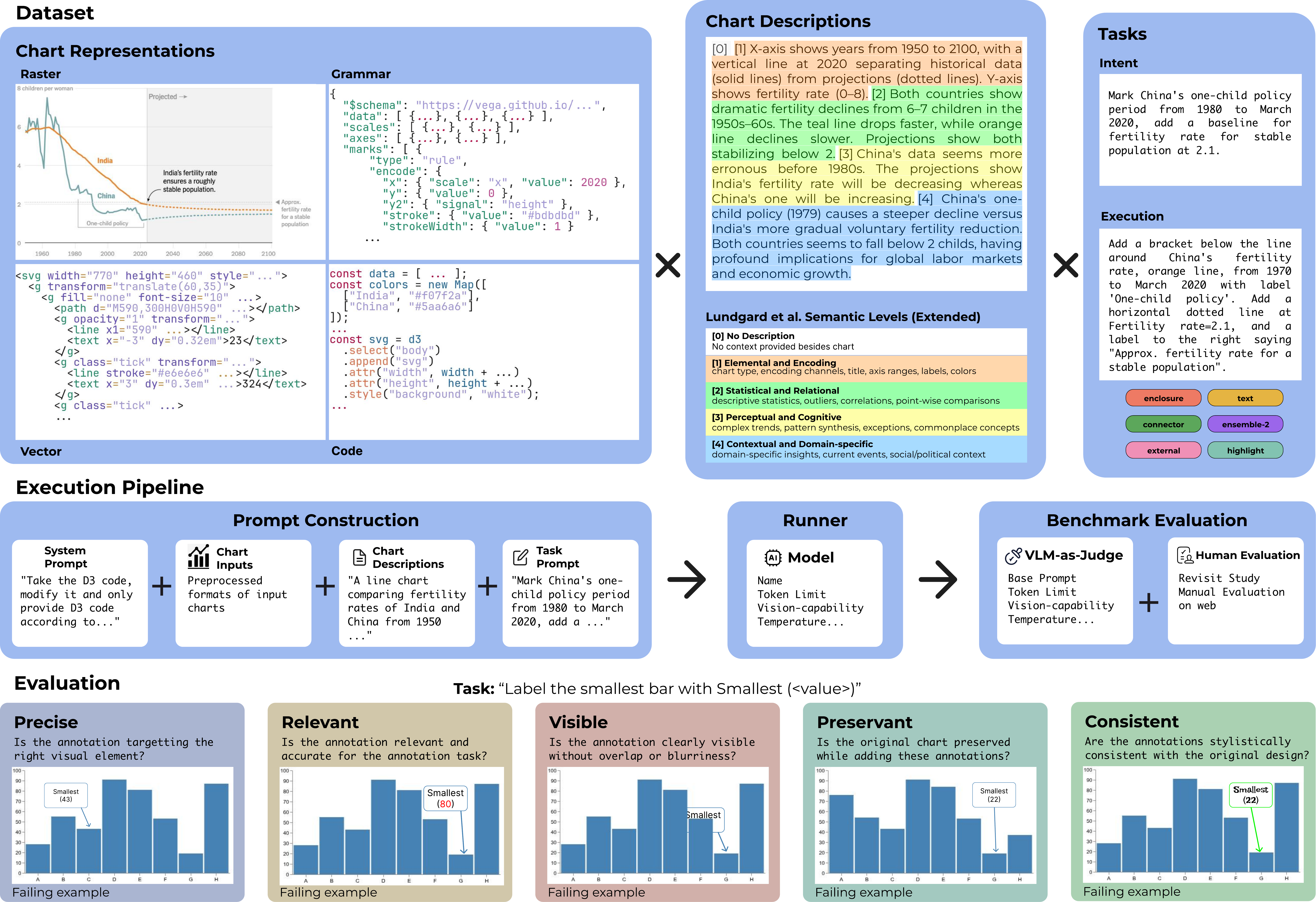}
    \caption{Overview of the AnnoBench dataset (top), execution pipeline (middle), and evaluation criteria (bottom).
    The top row illustrates the dataset design, where each chart has multiple representations (raster, vector, grammar, and code) and is paired with multi-level semantic captions ranging from no description to domain-specific context. Each chart is also associated with multiple annotation tasks defined at both the intent and execution levels. 
    The middle row illustrates AnnoBench's automatic prompt construction, running with the constructed prompt and providing environments for human and VLM-as-Judge evaluation. 
    The bottom row shows our five-dimensional evaluation framework, with a failing example in each criterion.
    }
    \label{fig:execution-pipeline}
    \vspace{-2em}
\end{figure*}

\autoref{sec:related-work} identifies both the factors a benchmark must control to support valid annotation experiments and the properties that matter for evaluating whether a generated annotation is correct. We organize AnnoBench around these two observations: benchmark design dimensions that control the conditions under which annotation is attempted (\autoref{sec:dataset-dimensions}), and an evaluation framework that defines what a correct annotation must satisfy under those conditions (\autoref{sec:eval-framework}).

\subsection{Dataset Design Dimensions}
\label{sec:dataset-dimensions}

Prior benchmark work discussed in \autoref{sec:related-work:llms} on chart understanding, captioning, and editing points to three factors that are especially consequential for annotation generation: chart representation, semantic context, and task specification. These factors correspond to three capabilities annotation requires simultaneously---identifying the intended target, determining what the chart supports, and integrating the requested addition into the visualization. AnnoBench treats each as an independent design dimension, so that failures can be attributed to the specific capability they stress: chart reading, semantic grounding, or annotation execution.

\parahead{Multiple Chart Representations}
Annotations, being a supplementary design element, are inherently bound to the form of the input they modify. 
Different representations change how difficult each of these steps becomes.

Raster inputs preserve rendered appearance and support visual reasoning, but they do not expose chart components as directly editable elements. Marks, axes, and labels must instead be inferred from appearance, and any modifications must be expressed at the pixel level, which may be beneficial or detrimental to models. Structured representations, such as vector markup, grammar-based specifications, and imperative code, expose chart components more explicitly and support direct manipulation, but require the model to reason over abstract structure rather than rendered appearance. A model may therefore succeed under one representation and fail under another for reasons that are separate from its underlying annotation performance.

This distinction matters for evaluation. A failure on a raster input may reflect weak visual localization, difficulty producing pixel-level modifications, or both. A failure on a structured representation may instead reflect difficulty parsing the format or producing well-formed edits, even when the required annotation is understood.
AnnoBench, therefore, treats chart representation as a controlled design dimension allowing the benchmark to distinguish failures in target identification, structural reasoning, and editing, and to attribute those failures more clearly to the criteria defined in \autoref{sec:eval-framework}.

\parahead{Task Specification}
The second axis of variation in AnnoBench is how the annotation goal is specified. The same objective can be expressed at different levels of detail---from a loose communicative intent to a precise specification of what to add, where to place it, and how it should connect to chart elements. Because prompt wording directly determines how much freedom a model has in deciding what to annotate and how, leaving this variation uncontrolled conflates failures of annotation strategy with failures of instruction following. AnnoBench makes it an explicit variable instead.

A task pairs a chart input with an annotation objective: the model receives one or more chart representations and is asked to produce the chart with the specified annotation integrated. A single chart may be paired with multiple tasks, each targeting a different annotation objective. Each task is defined at two prompt levels: \textbf{intent level} prompts describe the annotation goal without specifying its form or placement; \textbf{execution level} prompts specify what to add, where to place it, and how it should relate to chart elements.

This distinction separates two failure modes. A model that fails at the intent level cannot map a communicative goal to an appropriate annotation strategy. A model that succeeds only under execution-level prompts can follow explicit instructions but cannot reason about annotation strategy independently. Failures at both levels point to deeper problems---weak chart understanding, incorrect target identification, or unreliable placement decisions.

We draw tasks from the annotation taxonomy of Rahman~\etal{}~\cite{rahman2023qualitative}, which characterizes annotations along three dimensions: \emph{why} a given annotation type is used, \emph{what} data it targets, and \emph{how} it connects to chart elements and other annotations. Grounding tasks in this taxonomy connects AnnoBench to annotation practices observed in real-world visualizations rather than an ad hoc selection of task types. The specific task distribution and taxonomy coverage are described in~\autoref{sec:dataset}.

\parahead{Multi-level Semantic Descriptions}
How much a model knows about a chart directly shapes its annotation output~\cite{tang2023vistext}. Task accuracy depends not only on whether a model can read the chart visually, but on whether it understands what the chart is about. A model given only a chart must infer the chart's subject, domain, and key features from visual appearance alone.
However, a model given a rich textual description (such as encoding details about axis, marks, and channels, and other relevant context like trends and perceptual details) has that context explicitly available. 
Therefore, a failure in task accuracy might be mitigated by additional information, such as a caption or external domain information regarding the chart. Hence, we use the chart description semantics as a dimension of annotation capability. 

Each chart in AnnoBench is paired with multiple levels of semantic detail, applied independently of the selected representation type. 
We follow the semantic hierarchy of Lundgard~\etal{}~\cite{lundgard2022accessible}: level 1 covers chart type, encodings, and axis ranges; level 2 adds descriptive statistics and point-wise comparisons; level 3 adds perceptual patterns and trend synthesis; and level 4 adds domain-specific interpretation. The Lundgard framework was designed for accessibility, in which text entirely substitutes for the chart; here, descriptions augment the chart input rather than replace it. Therefore, we introduce level 0 (\emph{No Description}), in which we omit any additional context and serve the chart as a no-context baseline. 
All five levels are summarized in~\autoref{fig:execution-pipeline} along with an example.

These design dimensions define the conditions under which annotation is attempted. To evaluate model outputs under those conditions, we next define what a correct annotation must satisfy.

%% file: 04-eval-framework.tex
\subsection{Annotation Evaluation Framework}
\label{sec:eval-framework}
Evaluating annotation correctness requires per-instance conditions that can be checked directly: whether the annotation targets the intended visual element, asserts only what the underlying encoding supports, and fits within the chart layout without occluding evidence. As identified in \autoref{sec:related-work:annotations}, prior work does not define such conditions, leaving annotation quality without a basis for systematic comparison across systems. Because annotation tasks often admit multiple valid realizations~\cite{rahman2024qualitative, rahman2023exploring}, evaluating against a single reference output is also insufficient. A single aggregate score would conflate two fundamentally different failure types: an annotation that targets the wrong element, and one that is correct but styled differently than the reference. We developed the framework by examining what prior work on annotation systems~\cite{ren2017chartaccent, hullman2013contextifier}, chart overlays~\cite{kong2012graphical}, annotation grammars~\cite{rahman2025annogram, chen2025chartmark}, and annotation practice~\cite{rahman2025survey, rahman2024qualitative} implies about correctness. That work characterizes annotation through target binding, claim scope, and spatial placement, but as design dimensions rather than testable conditions. Through iterative discussion among the authors, we examined how each property applies when the question shifts from design to evaluation, yielding a set of criteria that apply consistently across annotation types, representations, and authoring contexts.

\parahead{Target Correctness (\preciseTag)}
An annotation makes a claim about a specific visual element or the 
chart as a whole, and that claim is only correct if the annotation 
is attached to the intended element~\cite{kong2012graphical, 
rahman2024qualitative}. For target-bound annotations, the 
annotation must refer to the correct mark, series, or region. 
Protocols that check only annotation text without verifying 
attachment cannot detect this class of error. In the failing 
example in \autoref{fig:execution-pipeline}, the task asks for 
the smallest bar to be labeled ``Smallest (<value>)''. 
The annotation produces the correct label text but attaches it to one of the 
taller bars instead. The content is plausible, but the target is 
wrong. Chart-level annotations do not refer to a specific visual target. For these, target correctness is satisfied by construction, and evaluation shifts to task accuracy. We operationalize this criterion as the rubric dimension \textbf{\precise{}}.

\parahead{Task Accuracy (\relevantTag)}
Correct target binding does not guarantee a correct annotation: an
annotation may be attached to the intended element while still
making an inaccurate or irrelevant claim about it. The \relevant{}
failing example in \autoref{fig:execution-pipeline} shows this
clearly: the label is attached to the correct smallest bar, but the
text is wrong---``Smallest (80)'' instead of ``Smallest (22)''. The
target is correct, but the claim is not. Task accuracy asks whether
the annotation makes an accurate and relevant claim about its
target. For target-bound annotations, the claim must be accurate
with respect to the specific mark, series, or region. For chart-level
annotations, it must be consistent with the visualization as a
whole. We evaluate this criterion independently of communicative
properties such as phrasing and framing, which remain
preference-sensitive. We use \textbf{\relevant{}} as the
corresponding rubric dimension.

\parahead{Spatial Feasibility (\visibleTag)}
An annotation must fit within the chart without obscuring evidence 
or creating ambiguity about what it refers to. It fails this 
condition if it overlaps axis labels, data marks, or other 
annotations, or if it is placed far enough from its target that 
the connection becomes unclear~\cite{kong2012graphical}. As shown 
in the \visible{} example in \autoref{fig:execution-pipeline}, the 
label ``Smallest'' is placed outside the chart area and connected 
to the correct bar only by a long line, making the association 
unclear. The annotation is neither occluding evidence nor 
incorrectly labeled---it simply cannot be unambiguously read as 
belonging to its target. Chart-level annotations face a weaker 
version of this condition: they must be placeable without 
obscuring evidence, but do not require proximity to a specific 
mark. We define this criterion as \textbf{\visible{}} in the 
evaluation rubric.

\parahead{Chart Preservation (\preservantTag)}
The underlying chart must remain faithful to its original encoding 
after the annotation is applied. Any automated system that generates 
or modifies a chart representation as part of an annotation---whether 
through code generation, SVG editing, or raster regeneration---risks 
inadvertently altering data encodings, axis scales, or mark 
positions in the process. For LLM-based systems specifically, this risk is compounded by the tendency of generative models to produce 
plausible-looking but factually incorrect content~\cite{huang2025survey}, which can potentially
manifest as modified bar heights, shifted axis ranges, or reordered 
categories in the output chart. The \preservant{} failing example 
in \autoref{fig:execution-pipeline} captures this well: the label 
``Smallest (22)'' targets the correct bar with the correct value, but the bar heights in the output chart have shifted relative to 
the original, changing the data the visualization encodes. The annotation itself is correct; the chart it was added to is not. 
Chart preservation evaluates whether the annotated output maintains 
the original visualization's data, structure, and visual mappings 
independent of the added annotation. We operationalize this as 
\textbf{\preservant{}} in the evaluation rubric.

\parahead{Stylistic Coherence (\consistantTag)}
An annotation should remain visually coherent with the chart it 
augments, including its typography, color use, and overall visual 
style~\cite{ren2017chartaccent}. Consider the \consistent{} 
failing example in \autoref{fig:execution-pipeline}: the label 
``Smallest (22)'' targets the correct bar and states the correct 
value, but is rendered in a large, bold style that clashes with 
the chart's minimal design. The annotation is factually correct 
and spatially feasible, yet visually discordant. Stylistic 
coherence is preference-sensitive: multiple realizations may be 
valid, so it is evaluated holistically rather than against a 
single reference. We use \textbf{\consistent{}} as the 
corresponding rubric dimension.

For each generated output, the evaluator receives the input chart, 
the model output, and the task description.
The five-dimensional rubric supports both manual human evaluation and 
automated VLM-as-a-Judge scoring, as described in 
Sect.~\ref{sec:evaluation}. We further suggest scoring a gradesheet for each rubric item, which can be found in \ourhref{https://annobench.insane.casa/rubrics}{annobench/rubrics}.

%% file: 05-annobench.tex
\section{AnnoBench Benchmark Dataset}
\label{sec:dataset}

\begin{figure}[ht]
    \centering
    \includegraphics[width=\linewidth]{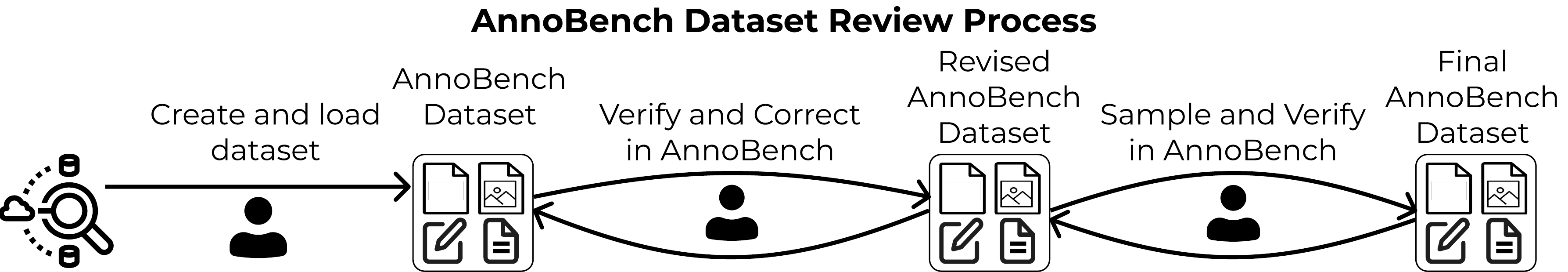}
    \caption{Overview of the AnnoBench dataset's iterative review process. Dataset items are constructed from online visualization galleries and journalism websites, then reviewed by two co-authors across chart representations, prompt levels, and multi-level chart descriptions. Finally, the same co-authors sampled benchmark items in the AnnoBench Browser and verified the final dataset entries.
    }
    \label{fig:dataset_review}
    \vspace{-0.5em}
\end{figure}

We now describe the construction of AnnoBench dataset. The dataset draws from two complementary sources: professionally produced charts from data journalism and charts collected from established visualization library galleries. This combination enables us to capture both the complexity and editorial nuance of real-world visualizations, as well as the controllability and reproducibility of library-based examples. 
For each chart included in the benchmark, we construct instances that vary along the benchmark dimensions introduced in \autoref{sec:design}, including chart representation, description condition, and task specification.
After constructing each dataset, we conducted an iterative review process to ensure data quality and consistency, shown in \autoref{fig:dataset_review}. 

\begin{table}[ht]
\centering
\resizebox{\linewidth}{!}{
\begin{tabular}{lccp{5cm}}
\toprule
\textbf{Dataset} & \textbf{\# Charts} & \textbf{\# Tasks} & \textbf{Chart Types} \\
\midrule

\ourhref{https://annobench.insane.casa/charts/professional-charts}{Professional Charts}
& 58 
& 58 
& Bar~(25), Line~(26), Scatter~(9)  \\

\midrule

\ourhref{https://annobench.insane.casa/charts/vega}{Vega Gallery}
& 92 
& 214 
& Scatter~(13), Heatmap~(9), Line~(9), Bar~(9), Others~(61)  \\

\ourhref{https://annobench.insane.casa/charts/vega-lite}{Vega-Lite Gallery}
& 192 
& 394 
& Line~(27), Bar~(24), Distribution~(18), Scatter~(13), Maps~(12), Others~(97)  \\

\bottomrule
\end{tabular}}
\caption{AnnoBench dataset combines professional charts with library-based examples, providing diverse chart types, multiple representations, varying description levels, and task prompt variants. 
}
\label{tab:dataset_summary}
\end{table}

\subsection{Professional Charts}
Our first dataset is a corpus of charts drawn from professional data journalism~\cite{rahman2023exploring}. We organize the corresponding annotation tasks using the taxonomy introduced by Rahman et al.~\cite{rahman2024qualitative}. The corpus draws from four newsroom sources with distinct visual and editorial conventions: 12 charts from \textit{The Washington Post}, 18 from \textit{The New York Times}, 15 from \textit{The Wall Street Journal}, and 15 from \textit{The Economist}. Because these visualizations were designed and published by professional teams at major news organizations, they reflect characteristics that make annotation difficult in practice, including editorial styling, custom typography, non-standard layouts, and dense labeling. Unlike the standardized outputs of popular visualization libraries, they also emerge from heterogeneous workflows involving tools such as Excel, Illustrator, Photoshop, proprietary newsroom pipelines, and custom in-house systems~\cite{parsons2021understanding}. Relative to the library-gallery examples (\autoref{sec:library-charts}), they introduce layout irregularities and design conventions that are uncommon in controlled gallery examples. The corpus consists of 25 bar charts, 26 line charts, and 9 scatterplots.

These charts also provide strong reference annotations for evaluation. Many already contain carefully designed highlights, callouts, labels, and explanatory overlays created by domain experts~\cite{rahman2023exploring}. These human-authored annotations reflect editorial intent and established practice. They serve as strong reference points, though a chart may admit other valid annotations.

Starting from the annotated source visualizations, we first create an unannotated version of each chart in D3.js, which serves as the benchmark's canonical source representation. From this D3 implementation, we derive additional representations, including SVG and grammar-based specifications, through a combination of manual conversion and LLM-assisted translation. Because many source articles do not provide the original data, we reconstruct or synthesize data that closely match the source visualization while preserving the perceptual and structural properties required for the annotation tasks.

We then derive benchmark tasks from the original annotated charts. In the professional corpus, each chart contributes a single base annotation objective corresponding to the published annotation. From this objective, we construct both intent-level and execution-level prompt variants and assign taxonomy-based metadata tags. Finally, we manually author five chart description conditions for each unannotated chart based on the source article. The full construction process is described in the supplementary materials. We make the final charts publicly accessible at \ourhref{https://annobench.insane.casa/charts}{annobench/charts}.

\subsection{Charts in Visualization Library Galleries}
\label{sec:library-charts}

To complement the professional charts dataset, we also collect charts from widely used visualization libraries. Gallery examples are common in chart understanding and visualization research~\cite{yang2024considering, kim2020answering, mcnutt2021integrated}, and they provide controlled examples with stable underlying specifications.

Although many visualization galleries could be included~\cite{yang2024considering}, supporting every ecosystem would require substantial engineering effort and evaluation cost. We therefore selected the Vega and Vega-Lite galleries because they provide diverse declarative chart specifications, stable source representations, and broad coverage of chart types while keeping the benchmark practical to build and maintain~\cite{satyanarayan2016vega, vega, yang2024considering}.

Consistent with the chart-representation dimension described in \autoref{sec:design}, these sources provide canonical code-based and grammar-based specifications, from which we also derive raster and vector representations.
Across the collected visualizations from Vega and Vega-Lite galleries, our library corpus covers a broad range of chart types, including bar, line, area, scatterplot, heatmap, histogram, map, node-link, radial, pie, and faceted chart variants (see~\autoref{tab:dataset_summary}). We also authored description conditions for the library charts. However, many gallery examples are designed primarily as usage demonstrations rather than domain-grounded visualizations, making it difficult to support richer contextual descriptions. As a result, the library corpus includes only the lower-level description conditions.

To ensure compatibility with model input constraints, we kept chart data inline where possible. For very large representations exceeding 16,384 tokens, we externalized the data and included a minimal example in the prompt to communicate the data schema. We also retained a small number of large inline examples to test long-context behavior.

We adopt a semi-automated task generation pipeline grounded in the annotation taxonomy of Rahman~\etal{}~\cite{rahman2024qualitative}. Instead of authoring all tasks manually, we designed three generation templates for each annotation type and used them to guide LLM-based task generation. To preserve a realistic distribution of annotation types, we used stratified sampling~\cite{neyman1992two} using the empirical distribution reported by Rahman~\etal{}. For each chart, we randomly assign sampled annotation types and prompt templates, then repeat this process four times, yielding four candidate tasks per chart. During generation, we provide the LLM with the chart representations and available description conditions to ensure the tasks remain consistent with the chart content. We then manually review the generated tasks, remove ambiguous or incompatible cases, and rewrite the remaining tasks as paired intent-level and execution-level prompts corresponding to the same annotation objective.

The Vega gallery contains 92 charts after excluding one example that demonstrates a video game; from these charts, we manually derive 214 annotation tasks. From the Vega-Lite gallery, after removing duplicates and filtering out small demonstration-only examples, we retain 192 charts and derive 394 annotation tasks.

AnnoBench datasets are designed to be extensible. Users can incorporate new datasets by following the standardized dataset schema without requiring additional code changes. To facilitate this, we also provide a curated set of charts from the D3 Observable Gallery, including multi-level semantic descriptions and multiple chart representations (D3, SVG, and raster). These charts can be directly explored in \ourhref{https://annobench.insane.casa/playground}{annobench/playground}, where users can experiment with custom prompts and annotation tasks. This dataset curation process is discussed in the supplementary document.

%% file: 06-annobench-system.tex
\section{AnnoBench Browser}
\label{sec:annobench_browser}

AnnoBench is designed as a multi-dimensional benchmark that captures the complexity of chart annotation across representations, semantic contexts, and task formulations. To fully leverage this richness, researchers must be able to systematically explore, configure, and analyze experiments across these dimensions. We therefore introduce the \emph{AnnoBench Browser}, an interactive web-based interface that operationalizes the benchmark as an executable and navigable system. The browser, as shown in~\autoref{fig:teaser}, enables users to explore datasets, configure experiments, and analyze model behavior within a unified framework, allowing them to study annotation performance across diverse conditions in a structured, interpretable manner.

The browser exposes AnnoBench as a structured design space, allowing users to select datasets, tasks, chart representations, semantic description levels, and prompt variants to define controlled experimental conditions. Given the diversity of representations and output formats in the benchmark, the system automatically generates prompts and standardizes model outputs, enabling direct comparisons across configurations and models.

A key feature of the browser is its ability to render and inspect model outputs across heterogeneous representations. For vector, grammar, and code-based charts, the system provides automatic compilation, visualization and error reporting, allowing users to evaluate both the rendered chart and its underlying representation.

The browser also supports simultaneous human and VLM-based evaluation, result aggregation across experimental conditions, and analysis of performance trends and evaluator agreement. The AnnoBench browser execution pipeline automatically compiles supported chart representations into static images and uses VLM-Judge for automated evaluation.
To support larger-scale evaluation, the system allows exporting results to ReVISit~\cite{2025_vis_revisit} studies, enabling standardized human evaluation workflows or crowdsourcing. The system can further import study results from ReVISit back into the AnnoBench browser for unified analysis.
Together, these capabilities enable reproducible, large-scale benchmarking and fine-grained analysis of annotation behavior. Additional implementation details and system capabilities are described in the supplementary document.

%% file: 07-benchmark-evaluation.tex
\section{Experiments}
\label{sec:evaluation}

We describe a series of experiments designed to evaluate the dimensions of the AnnoBench benchmark through controlled studies with contemporary models. Our goal is to understand how different design dimensions of AnnoBench influence annotation performance, rather than solely benchmarking models. Specifically, we conduct experiments that isolate and analyze the effects of chart representation~(\cref{sec:representation_study}), task-prompt specificity~(\cref{sec:intent_execution_study}), and semantic chart descriptions~(\cref{sec:description_study}), enabling us to assess how each dimension contributes to the difficulty and variability of chart annotation tasks.

A key challenge is the scale of the benchmark, as the combinatorial space of configurations makes exhaustive evaluation computationally and practically prohibitive.
Even restricting ourselves to the core dimensions in this paper would produce more than 273,960 output charts to generate and assess. Assuming an average of 4,000 input tokens and 4,000 output tokens per task, running this full matrix using the March 2026 Claude Opus 4.6 API rates would cost approximately \$98,625 (\$15/MTok input and \$75/MTok output)~\cite{AnthropicPricing2026}, before accounting for retries, rendering failures, and human evaluation effort. The resulting manual assessment burden would be 2,283 hours (assuming 30 seconds per chart), which is even less practical.

In place of such an enumerative evaluation, we conduct four one-dimension-at-a-time experiments. We conduct four experiments evaluating different dimensions. First, we look at the effect of representation, then prompt form, the effect of captions, and lastly model performance. 
In each experiment, we use what are believed to be the best configurations for each dimension, except when that dimension is under test. For instance, we use execution level prompting in all experiments besides the second, while in the first experiment, we use the most performant representation (which is then used in other experiments). 
This strategy keeps human and API costs
manageable while still supporting controlled analysis of how
individual dimensions, such as representation or prompt specificity,
affect performance.

\parahead{Human and VLM-based Evaluation}
For the first three experiments, we use stratified sampling~\cite{neyman1992two} with $n=15$ tasks per experiment condition. 
To increase sample size and improve statistical accuracy, we use AnnoBench's VLM-as-a-Judge feature to score an additional sample of equal size ($n=30$). Because each study multiplies this base sample by the number of compared levels in its independent variable and by the number of tested models, the final number of scored outputs remains tractable while still supporting controlled comparisons. Under this design, the first experiment contains $15 \times 4 \times 5 = 300$ samples, the second contains $15 \times 3 \times 5 = 225$ samples, and the third contains $15 \times 2 \times 5 = 150$ samples.

All samples in these experiments are scored by two authors and three
VLM judges using the five-dimensional rubric introduced
in~\autoref{sec:eval-framework}. The three VLM judges are GPT-5.2
(\texttt{gpt-5.2}), Claude Sonnet 4
(\texttt{claude-sonnet-4-20250514}), and Gemini 2.5 Flash
(\texttt{gemini-2.5-flash}).

The generated annotations are produced by five generation models under
evaluation: GPT-5.2, Claude Sonnet 4, Gemini 2.5 Flash, DeepSeek-R1 14B
(\texttt{deepseek-r1:14b}), and Gemma 3 27B
(\texttt{gemma3:27b}). Thus, GPT-5.2, Claude Sonnet 4, and Gemini 2.5
Flash serve as both generation models and VLM judges, whereas
DeepSeek-R1 14B and Gemma 3 27B serve only as generation models.

We report inter-rater reliability (IRR) for the human evaluators in each
experiment to show consistency in manual scoring. As noted
in~\autoref{sec:related-work}, prior work does not yet establish how
reliably VLM-as-a-Judge evaluation can assess visualization annotations.
We therefore analyze the agreement between VLM-based and human ratings
and discuss where automated judging is reliable and where it fails at
the end of~\autoref{sec:representation_study}.

The AnnoBench analysis interface supports both manual filtering of runs and configuration-based selection of experiments. For any selected configuration, the interface presents aggregate scores, IRR between human evaluators, agreement between human and VLM-based judgments, and summary visualizations of model performance across dimensions and rubric categories. This analysis workflow is important not only for reporting final results, but also for understanding where annotation failures arise. All model run results, along with their human and VLM-judge evaluation results, are explorable in~\ourhref{https://annobench.insane.casa/analysis}{annobench/analysis}.

\parahead{Sampling Strategy}
To avoid bias toward any single annotation type, we use proportionate stratified sampling over annotation categories (\eg{} text, highlight, enclosure) for all experiments, following the empirical distribution observed in our corpus and motivated by prior studies of real-world annotation usage~\cite{rahman2023qualitative}. This yields evaluation sets that better reflect the distribution of annotation practices found in the wild.

\parahead{Failure Handling}
For non-raster chart representations, models generate text that is compiled to the final annotated chart.
While contemporary models are effective at generating text, they sometimes generate invalid or malformed specifications that do not run. 
When a model output fails, the system can provide execution logs feedback and re-invoke the model to obtain a corrected output. 
This mechanism helps separate failures of \emph{annotation reasoning} from failures of \emph{format compliance}. We allow up to three automated retry attempts per failed generation to correct these incidental errors to avoid an open-ended repair loop.
If the output remains non-executable, we assign a score of 0 for that trial across all evaluation categories. This policy is intentionally conservative: a non-renderable output is unusable in practice.

\subsection{\texttt{Experiment 1:} Evaluating Representation Robustness}
\label{sec:representation_study}

Our first experiment evaluates the effect of chart representations. Different representations expose different kinds of information to the model. For instance, while raster images preserve visual appearance, they hide structural editability. Similarly, 
code-level representations expose the full construction process, with a lower level of abstraction that may not be aware of the semantics of the final image. 
This experiment sought to answer the question---\emph{Which chart representations best support annotation generation?} We evaluate models across four core representation formats: \textbf{Raster} (provided as base64-encoded image content or image-grid input, depending on model API support), \textbf{Vector} (SVG), \textbf{Grammar} (Vega specification), and \textbf{Code} (JS with D3 implementation). The underlying tasks are matched across other conditions, allowing us to isolate the effect to only the representation dimension.

\begin{figure*}[t]
    \centering
    \includegraphics[width=\linewidth]{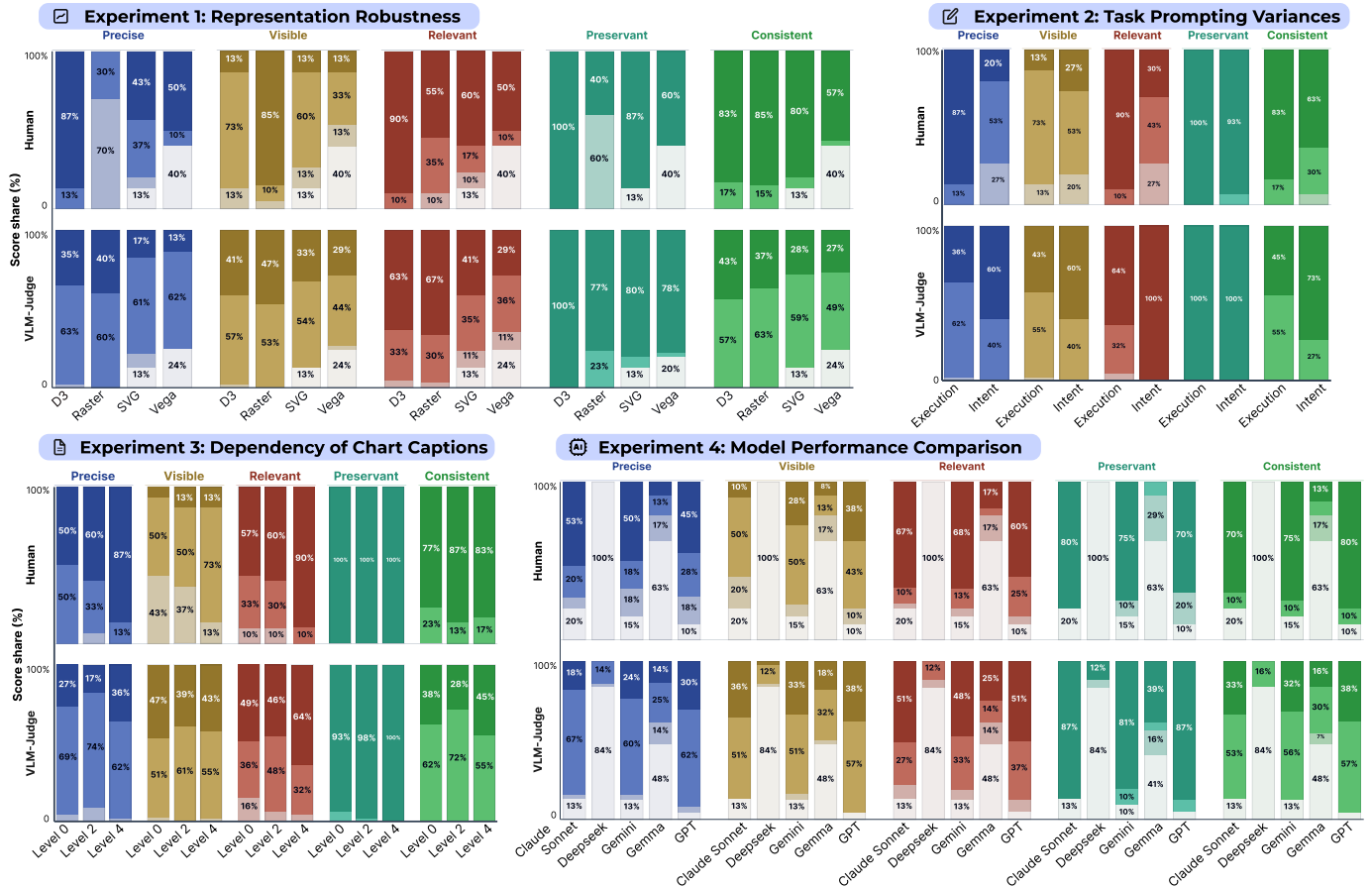}
    \caption{Distribution of rubric scores for Experiment~1--4. In each chart of the experiments, each column group corresponds to each evaluation criterion, and rows correspond to the evaluator (Human or VLM-Judge). Each column corresponds to the dimension we are evaluating in the experiment that it represents. Within each bar, score percentages are shown between 0 (lowest intensity) and 3 (highest intensity).
    }
    \vspace{-4pt}
    \label{fig:result-barcharts}
\end{figure*}

\parahead{Results}
\autoref{fig:result-barcharts} (Experiment 1) summarizes the score
distributions across the evaluation rubric using the average scores of
two human visualization experts and three VLM judges, applied to outputs
from five generation models under evaluation: GPT-5.2, Claude Sonnet 4,
Gemini 2.5 Flash, DeepSeek-R1 14B, and Gemma 3 27B.

Even with the three-retry-and-fix prompting process, we find
DeepSeek-R1 14B to have compilation errors or completely missing
annotations in all outputs, scoring zero across all evaluation criteria.
Looking at the examples, we see the model continuously providing
instructions for adding annotations to the chart, rather than returning
the completed chart for each retry. Next, Claude Sonnet 4 had compilation errors in 18\% of benchmark runs,
Gemini 2.5 Flash in 13\%, and GPT-5.2 in 8\%. Among these errors from
all models, approximately 30\% were from SVG, 70\% were from Vega
specifications, and none were caused by D3 charts. Due to the poor performance of DeepSeek-R1 14B, we exclude it from the
remaining experiments.

Across the two human evaluators, Cohen's $\kappa$ exceeds 0.93 for all five rubric dimensions (\precise{}, \visible{}, \preservant{}, \relevant{}, and \consistent{}), indicating near-perfect agreement. We further compute the mean absolute distance (MAD) between the two human evaluators and find it to be below 0.22 for all rubric dimensions (scored 0-3), suggesting that disagreements are not only infrequent but also small in magnitude.

For human evaluation results, the D3 code representation achieves the highest scores across all dimensions, except \visible{}, where raster performs better. 
The Vega representation followed D3 on \visible{}, \relevant{}, and \preservant{}, while SVG performed better on \precise{} and \consistent{} than Vega.
We adopt the D3 code representation for the remaining experiments as the benchmark's default best-effort representation.

The VLM-as-a-Judge results show a broadly similar ordering among D3,
SVG, and Vega representations ($\rho > 0.4$ for all, with the Gemini
2.5 Flash judge yielding the highest agreement), but much weaker
agreement with human evaluation for raster outputs (average $\rho = 0.3$). Across VLM judges, raster appears to
perform slightly better than D3. To examine this disagreement, we
computed human--VLM correlations for each chart representation and
rubric category (see~\autoref{tab:human-vlm-correlation}). For raster
outputs, agreement is weak, with $\rho < 0.4$ in most rubric
categories. We discuss qualitative failure cases and possible reasons
in~\autoref{sec:discussion}. We also separately tested vector, grammar, and code representations for a chart with large inline data. As expected, all models returned
truncated outputs in all three cases, yielding zero scores across all
evaluation dimensions.

This experiment yields three important findings. First, chart annotation capability varies across representation types, with code-based inputs performing best overall in human evaluation. Second, raster chart annotation remains a major weakness for current models: although models can often generate visible annotation-like elements, they often fail to faithfully preserve the underlying charts. Third, VLM-as-a-Judge evaluation is itself representation-sensitive. While it is moderately reliable for structured textual chart outputs, it is much less reliable for raster outputs, where it tends to overestimate quality.

\definecolor{colFair}{HTML}{e66101}        %
\definecolor{colModerate}{HTML}{fdb863}    %
\definecolor{colSubstantial}{HTML}{b2abd2} %
\definecolor{colPerfect}{HTML}{5e3c99}     %

\newcommand{\autoscore}[1]{%
\begingroup
\edef\temp{#1}%
\ifdim \temp pt < 0.4pt
\textcolor{colFair}{\textbf{#1}}%
\else\ifdim \temp pt < 0.6pt
\textcolor{colModerate}{\textbf{#1}}%
\else\ifdim \temp pt < 0.8pt
\textcolor{colSubstantial}{\textbf{#1}}%
\else
\textcolor{colPerfect}{\textbf{#1}}%
\fi\fi\fi
\endgroup
}

\setlength{\tabcolsep}{4pt} %
\begin{table}[!ht]
\centering
\caption{Spearman correlation ($\rho$) between human evaluation and the best-performing VLM judge for each chart representation (Numeral colors mapped $\rho \in [0,1]$ (red--green)). Agreement is weakest for raster charts and strongest for vector-based outputs.}
\label{tab:human-vlm-correlation}
\resizebox{\linewidth}{!}{
\begin{tabular}{
p{1.2cm}
>{\centering\arraybackslash}p{0.9cm}
>{\centering\arraybackslash}p{1cm}
>{\centering\arraybackslash}p{1.1cm}
>{\centering\arraybackslash}p{1.3cm}
>{\centering\arraybackslash}p{1.2cm}
}

\toprule
 & \textbf{Precise} & \textbf{Visible} & \textbf{Relevant} & \textbf{Preservant} & \textbf{Consistent} \\
\midrule
D3 &
\autoscore{0.63} &
\autoscore{0.44} &
\autoscore{0.54} &
\autoscore{0.72} &
\autoscore{0.41} \\

Grammar &
\autoscore{0.64} &
\autoscore{0.48} &
\autoscore{0.73} &
\autoscore{0.57} &
\autoscore{0.54} \\

Vector &
\autoscore{0.75} &
\autoscore{0.80} &
\autoscore{0.79} &
\autoscore{0.97} &
\autoscore{0.84} \\

Raster &
\autoscore{0.32} &
\autoscore{0.22} &
\autoscore{0.48} &
\autoscore{0.39} &
\autoscore{0.47} \\
\bottomrule
\end{tabular}
}
{\tiny 
\fcolorbox{black}{colFair}{\rule{0pt}{1pt}\rule{1pt}{0pt}} \raisebox{-1pt}{fair (0.2--0.4)} \ 
\fcolorbox{black}{colModerate}{\rule{0pt}{1pt}\rule{1pt}{0pt}} \raisebox{-1pt}{moderate (0.4--0.6)} \ 
\fcolorbox{black}{colSubstantial}{\rule{0pt}{1pt}\rule{1pt}{0pt}} \raisebox{-1pt}{substantial (0.6--0.8)} \ 
\fcolorbox{black}{colPerfect}{\rule{0pt}{1pt}\rule{1pt}{0pt}} \raisebox{-1pt}{almost perfect (0.8--1.0)}
}
\end{table}

\subsection{\texttt{Experiment 2:} Evaluating Task Prompting Variances}
\label{sec:intent_execution_study}

Our second experiment investigates the effects of task specificity. A given annotation goal can be framed either as a high-level intent or as an explicit procedural instruction (see tasks in~\autoref{fig:execution-pipeline}). 
To this end, we sought to understand---\emph{How does the level of prompt specificity influence the annotation capability of models?}

We draw on a similar design to the previous experiment, but with representation fixed and task form varied using the best-performing chart representation (D3).
We consider both prompt levels present in the data set, intent-level (which keep task open-ended) and execution-level (which explicitly constrains the annotation type)
This experiment probes whether models can independently infer effective annotation strategies or primarily succeed when given concrete guidance.

\parahead{Results}
\autoref{fig:result-barcharts} (Experiment 2) shows the score distributions for human and VLM-as-Judge evaluators.
We find that the execution-level prompt achieves higher average performance across all evaluation criteria for human-evaluator scores, which aligns with our expectations because explicit instructions reduce ambiguity and constrain the solution space to our expectations for the annotation task.

Inter-rater reliability between the two human evaluators was high. They had near-perfect agreement (exact score matches for \precise{}, \preservant{}, and \visible{}) across most samples. For \consistent{}, however, agreement was lower (average Spearman $\rho \approx 0.6$), reflecting the inherently subjective nature of stylistic evaluation. Judging styling consistency can be highly subjective, especially with the open-ended nature of intent-level prompts.

We find the improvements to be most pronounced in \precise{}, \preservant{}, and \visible{}, suggesting that explicit instructions help models correctly localize annotations, preserve the input chart, and place annotations without overlap or conflict with other chart elements.

Since this experiment is performed only with the code representation, where models tend to perform best, most annotation tasks received comparably high scores from both humans and VLM judges across all evaluation factors. However, VLM-Judge tends to score slightly lower on tasks with execution-level prompts and higher on those with intent-level prompts. 
As a result, intent-level prompts receive slightly higher scores by VLM-judges than execution-level prompts. We suspect that this discrepancy arises from the evaluation behavior of VLM judges. Because intent-level prompts are less constrained, it is easier for a judge model to rationalize that an output satisfies the prompt's intent, even when the annotation may look suboptimal to a human evaluator. In other words, VLM judges tend to reward semantic plausibility over structural correctness.

We notice a key behavioral difference across prompt levels in the selection of annotation types. For example, when asked to ``highlight the significant wave duration for Delta wave...'' models operating under intent-level prompts typically use simple strategies such as adding a text label or arrow. In contrast, execution-level prompts that explicitly specify an enclosure (\eg{} ``add a vertical shaded y-spanning region during the Delta wave...'') consistently produce the intended enclosure-based annotation (see third chart in~\autoref{fig:teaser} (B)). Notably, we find that intent-level prompts almost never result in enclosure annotations unless explicitly requested.

These results highlight a limitation in current models: although they can interpret high-level goals, they often default to simpler or more common annotation strategies (\eg{} text, connector, and indicator~\cite{rahman2024qualitative}) rather than exploring the full design space of possible annotations. This pattern is consistent with broader observations in LLM research, where models are less reliable under underspecified prompts and remain weaker at autonomous planning than when task requirements are made explicit~\cite{yang2025prompts,valmeekam2023planning}.

\subsection{\texttt{Experiment 3:} Evaluating Dependency of Chart Captions}
\label{sec:description_study}

Our third experiment evaluates how performance depends on the accompanying textual chart descriptions. 
Although AnnoBench includes five semantic caption levels, it is not clear whether models genuinely benefit from richer chart descriptions once they are already provided with a complete chart, such as code or grammar representations. 
Here we sought to understand---\emph{How much do models rely on textual chart descriptions when generating annotations?}

We continue to follow the default evaluation setup, but vary the semantic description level across the lowest (No Description), mid (Statistical and Relational), and highest (Contextual and Domain-Specific, or the highest description available). 

\parahead{Results}
\autoref{fig:result-barcharts} (Experiment 3) shows score distributions across rubric dimensions for both human evaluators and the VLM-as-Judge. Overall, human evaluation shows a moderate improvement in \precise{}, \visible{}, and \relevant{} as the semantic description level increases. 
\preservant{} remains constant across all conditions, with perfect scores (100\%) for all description levels. This is expected as we have seen in~\autoref{sec:representation_study}, as preservation primarily depends on the chart representation (constant D3 code in this experiment) rather than semantic context. 
The \consistent{} scores vary across levels, suggesting that while semantic descriptions help with task alignment, stylistic consistency is influenced more by the prompt specificity and design guidelines (as seen in~\autoref{sec:intent_execution_study}).

The VLM-as-Judge results exhibit weaker sensitivity to semantic description levels compared to human evaluation. Across all evaluation criteria, score distributions remain relatively similar, with only minor variations. 
This behavior suggests that VLM judges rely more heavily on the final visual plausibility of the output rather than on how well the annotation aligns with the underlying chart semantics, since the VLM-Judges were not provided with the same chart semantic descriptions to avoid bias. As a result, improvements driven by richer semantic descriptions are less noticed by VLM evaluations.

This experiment shows that semantic descriptions provide measurable but secondary benefits for chart annotation. Richer descriptions
improve performance mainly in dimensions tied to semantic
understanding, such as \precise{} and \relevant{}, but have limited
effect on stylistic consistency and chart preservation.
Finally, semantic description augmentation cannot fully compensate for limitations in a model's ability of annotating visualizaiton charts, and improving robustness in these areas remains a key direction for future work.

\subsection{\texttt{Experiment 4:} Model Performance comparison}
\label{sec:model_comparison}
Our final experiment compares the overall annotation capability of state-of-the-art models under the standardized, best-effort configuration. 
While previous experiments isolate the effects of representation, prompt specificity, and semantic descriptions, it remains unclear how different models perform when these factors are optimized. 
Here we seek to understand---\emph{How do current state-of-the-art models compare in their ability to generate chart annotations.}

We use the same results from all previous experiments and compare each experimental setup, both aggregated and individually, to identify whether model performance varies across configurations, where one model performs better in one configuration but poorly in another.

\parahead{Results}
\autoref{fig:result-barcharts} (Experiment 4) shows the average distribution of scores across rubric dimensions for all evaluated models. Across human evaluation, we observe a clear separation between higher-performing and lower-performing models. 
GPT-5.2 and Claude Sonnet 4 consistently achieve the highest scores
across most evaluation criteria, with Gemini 2.5 Flash following
closely behind.

In contrast, smaller or less specialized models such as DeepSeek-R1 14B
and Gemma 3 27B perform substantially worse across all criteria.
Notably, DeepSeek-R1 14B frequently fails to produce valid or meaningful
annotations, resulting in near-zero scores across multiple dimensions.
Gemma 3 27B demonstrates partial capability but struggles with
\precise{} and \relevant{}, often producing incomplete or misaligned
annotations.

Across individual experimental conditions, we observe that the relative ranking of models remains largely consistent. Models that perform well in one configuration tend to perform well across others, indicating that improvements are not tied to a specific representation, prompt type, or semantic context. This stability suggests that annotation capability reflects a more general underlying competence rather than sensitivity to a particular dimension.

Finally, even the strongest models exhibit variability across evaluation dimensions, suggesting that chart annotation remains a challenging, multifaceted task requiring both semantic understanding and precise visual execution to add annotations to any chart representation.

%% file: 08-discussion.tex
\section{Discussion \& Conclusion}
\label{sec:discussion}
Chart annotation is a multi-dimensional challenge that depends jointly on representation, instruction specificity, and semantic context. No single factor alone determines success; rather, annotation quality emerges from the interaction between chart understanding and the ability to execute precise visual modifications, which we evaluated with a five-dimensional evaluation framework. While evaluating annotations from various models, we encountered some key observations.

\parahead{Chart representations have the largest impact}
Results from Experiment 1 show that D3 substantially outperforms other chart representations for annotation authoring, likely for two key factors: its JavaScript foundation is well represented in LLM training data, making it easier for models to manipulate, and annotations can be added incrementally without regenerating the entire chart. In contrast, raster representations frequently lead to severe \preservant{} failures, where even simple annotation attempts introduce hallucinated marks or distort the chart to the point of unusability; models may also shift data-encoded marks to create space for annotations, artificially inflating \visible{} scores (see~\autoref{fig:teaser} (B) and supplementary document for examples). SVG exhibits a different limitation: while it preserves structure, its verbosity often pushes models beyond context or even token limits, resulting in truncated, malformed, or hallucinated outputs. Overall, the performance gap across representations is striking---highlighting that structured representations play a critical role in enabling reliable annotation generation and should be prioritized in future visualization systems.

\parahead{Prompt specificity improves execution but exposes limited design inference}
Results from Experiment~2 show that models benefit substantially from explicit execution-level prompts, which consistently outperform intent-level prompts across most evaluation criteria. This suggests that current models are much better at \emph{following} annotation instructions than at \emph{inferring} an appropriate annotation strategy from a high-level communicative goal. In particular, intent-level prompts often lead models to default to annotation types that do not better match the intended message. As chart annotation is not only a matter of chart understanding, but also of visual planning and design reasoning, the results suggest that improving annotation quality will require models that can reason more flexibly about alternative annotation strategies, rather than relying primarily on explicit procedural guidance.

\parahead{Semantic chart descriptions help when annotations are relevant to it}
Experiment~3 shows that richer chart descriptions provide measurable, though secondary, benefits for annotation generation. Higher-level chart captions improve performance primarily on dimensions tied to semantic understanding, especially \precise{} and \relevant{}, indicating that models use textual context to better identify what should be annotated and why. However, these gains are modest compared with those from chart representation or prompt specificity, and they do not improve \preservant{}. This suggests that semantic descriptions can support chart interpretation, but they cannot compensate for weaknesses in structured editing, spatial reasoning, or visual design. In other words, giving a model more context about the chart helps it understand the task, but does not necessarily help it carry out the annotation well.

\parahead{VLM-based judges overestimate quality on raster outputs}
Qualitative analysis of disagreement cases reveals that VLM-based judges tend to over-credit raster outputs, often judging annotations as correctly placed relative to the original chart even when the underlying chart content has been distorted. In such cases, annotations appear visually plausible but no longer correspond to the correct data, leading to low human scores in \precise{}, \relevant{}, and \preservant{} (see supplementary document for examples). While automated judging aligns reasonably well with human evaluation for structured representations, it performs less reliably on raster inputs, where it struggles to detect input corruption and may overestimate semantically incorrect results. These findings suggest that, although VLM-based judging is useful for scaling evaluation, it should be applied with caution in settings involving high design flexibility, such as annotation tasks.

\parahead{Extensibility of AnnoBench Browser}
AnnoBench Browser is released as a Python library and is designed to be
extensible along several dimensions. Users can add new visualization datasets by following the documentation and standardized data schema. For unsupported chart representations, custom rendering functions can be registered through the API, enabling integration with additional visualization frameworks. Furthermore, the system supports extension of semantic description strategies and annotation prompt levels, allowing researchers to introduce new chart description methodologies or task variations. Importantly, AnnoBench is not limited to LLMs or VLMs; it can be used to evaluate any system that takes a chart and a task prompt as input and produces an annotated chart as output, including rule-based methods, program synthesis approaches, or hybrid human-AI workflows.

\parahead{Limitations}
Despite the breadth of AnnoBench, it has several limitations. 
First, although we consider multiple chart representations and datasets, our evaluation covers only a subset of possible configurations due to practical constraints on API cost and human assessment effort. We also evaluate a limited set of models; while sufficient for comparative analysis, broader model coverage would strengthen generality.
Second, while we include a diverse set of annotation types, our tasks are constrained by the distributions present in available datasets and may not fully capture all real-world annotation scenarios. Third, our evaluation varies one dimension at a time, which enables controlled analysis but may miss higher-order interactions between dimensions.
Fourth, our evaluation relies on a fixed rubric and a limited number of human evaluators. Although inter-rater reliability is high, some dimensions, particularly stylistic consistency, remain inherently subjective. While we incorporate VLM-based judging to scale evaluation, we observe clear failure modes, and such methods should not be treated as a substitute for human assessment. Designing more reliable automated judges for visual tasks remains an open challenge.
Fifth, our experiments focus on static charts and do not consider interactive or animated visualizations, which introduce additional complexities for both annotation design and evaluation.
Finally, model capabilities evolve rapidly, which may reduce the long-term relevance of the specific model performance results reported in this paper. For this reason, we do not include model outputs in the benchmark dataset and instead present them as a point-in-time evaluation to demonstrate the utility of AnnoBench for analyzing visualization annotation dimensions.

\parahead{Future work}
AnnoBench opens several directions for future research. From a dataset perspective, we plan to expand the benchmark by adding additional chart sources, including multi-level task prompts to existing datasets such as the D3 Observable Gallery (\autoref{sec:dataset}), as well as incorporating new visualization ecosystems, interactive and animated charts, and domain-specific datasets. Increasing the diversity of chart types, data domains, and annotation tasks will further strengthen its role as a general benchmark for understanding and editing visualization. 
Beyond dataset expansion, AnnoBench provides a foundation for studying visualization tasks more broadly. While this work focuses on chart annotation, the underlying framework supports any task that involves visualizations and produces either textual or visual outputs. We plan to extend AnnoBench into a unified platform for evaluating a wider class of visualization tasks, including chart editing, transformation, and explanation, and to explore hybrid evaluation approaches that combine rule-based validation with model-based judging to better capture the complexities of visualization tasks.

%% file: 09-conclusion.tex
\parahead{Conclusion}
We present AnnoBench, a benchmark for evaluating chart annotation through per-instance correctness criteria and controlled variation in chart representation, task specification, and semantic context. Across our experiments, current models show a clear gap between understanding what a chart says and correctly editing it to express that understanding. Performance improves substantially when models receive structured, editable chart representations and explicit execution-level instructions, but raster inputs, underspecified prompts, and chart-preservation failures remain persistent weaknesses. These findings suggest that progress in annotation automation will require more than stronger chart comprehension alone. It will also require methods that support faithful chart editing, stronger inference of annotation strategies, and evaluation procedures that do not reward visually plausible but semantically incorrect outputs. AnnoBench provides a concrete basis for that work.

%% file: 10-appendix.tex
\setcounter{page}{1}
\onecolumn
\renewcommand{\thesection}{S\arabic{section}} %
\setcounter{section}{0} %
\begin{center}
    {\LARGE\textbf{Appendix}}
\end{center}

\section{AnnoBench Browser Capabilities}

\parahead{Exploration and Dataset Navigation}
AnnoBench supports flexible filtering and selection of dataset charts and benchmark runs directly through the browser interface. Users can select models, datasets, task subsets, chart representations, semantic description levels, and prompt types to define custom experimental conditions. These selections can then be used for exploration, running with newer models, ablation tests, or any other purposes.

\parahead{Experiment Configuration and Execution}
Based on the subset selection, users can configure a new configuration for new experiments.
Because evaluating chart annotation often involves multiple representations and output formats, the system automatically handles prompt construction and result processing based on the selected configuration. For example, users may provide a single representation (\eg{} raster or D3 code) or combine multiple representations within the same prompt. The system adapts the prompt template and output expectations accordingly, enabling direct comparison between unimodal and multi-representation settings.

To support fast, practical, large-volume experimentation, the browser integrates with an asynchronous execution pipeline that can run multiple tasks concurrently. All experiment configurations, inputs, outputs, and metadata are stored in a standardized format, allowing results to be reproduced, shared, and re-analyzed.

\section{D3 Observable Gallery Dataset}
We also collect charts from the D3 Observable Gallery~\cite{observableD3Gallery}, which provides low-level code-based chart representations that serve as canonical source inputs. Because Observable examples are notebook-specific, we standardize them into a consistent JavaScript format suitable for benchmarking. We exclude 56 interactive charts and 24 animated charts that fall outside the scope of our study, and convert the remaining 53 static charts into standardized JavaScript representations. 
We retain this corpus as an auxiliary resource for chart representations and description conditions rather than as part of the task-bearing benchmark. It broadens the coverage of source representations available in AnnoBench and supports future extensions. AnnoBench is designed to be extensible, and we provide a clear dataset structure and instructions for adding additional datasets in the documentation.

\section{Implementation Details}

AnnoBench is implemented as a Python library with a web-based interface, following a data-driven design. Each dataset is organized into charts and tasks, where charts include multiple representations (e.g., raster images, SVG, Vega/Vega-Lite specifications, or D3 code), and tasks define annotation objectives, prompt variants, and evaluation settings. This structure allows the same chart to be presented in different modalities while keeping tasks consistent across representations.

The benchmark pipeline automatically constructs model inputs by combining the selected chart representations, semantic descriptions, and task prompts. Depending on the configuration, models may receive visual inputs, structured specifications, or code representations. Outputs are then standardized and, when necessary, rendered into images so that both humans and automated methods can evaluate the results consistently across representations.

AnnoBench supports multiple evaluation strategies, including rule-based checks, similarity-based comparisons, human evaluation, and VLM-as-a-judge methods. The system is designed to handle errors during generation or rendering and to allow rerunning failed cases, separating model reasoning failures from execution issues. 

To support large-scale experimentation, the system manages benchmark runs, stores results in a structured format, and enables reproducible analysis across different configurations. The web interface provides tools for launching experiments, inspecting model outputs, and analyzing results across datasets, models, and evaluation criteria. Overall, the implementation is designed to be flexible and extensible, allowing researchers to evaluate diverse models, chart representations, and annotation tasks within a unified framework.

\section{Dataset Creation Process}

\begin{figure}[ht]
    \centering
    \includegraphics[width=0.5\linewidth]{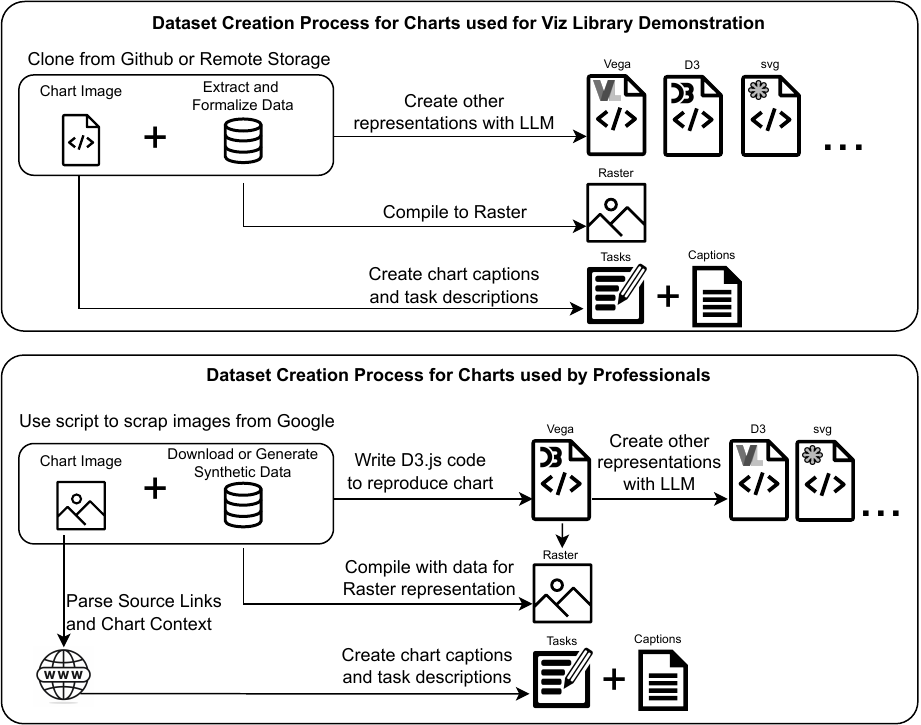}
    \caption{Overview of the AnnoBench dataset construction and review process. Top panel illustrates the pipeline for charts collected from visualization libraries, where chart images and data are extracted from repositories, additional representations (\eg{} Vega, D3, SVG) are translated with LLMs and manually verified, raster versions are compiled, and corresponding captions and annotation tasks are created. The bottom panel shows the pipeline for professionally produced charts, where images are scraped from online sources, data are reconstructed or synthesized, charts are reimplemented in D3.js, and additional representations are generated before producing raster outputs, captions, and tasks.
    }
    \label{fig:dataset-creation}
\end{figure}

\autoref{fig:dataset-creation} illustrates the dataset construction process for both visualization library charts and professionally produced charts. For library-based charts, we start from publicly available chart specifications or examples, normalize the underlying data, and generate multiple chart representations (e.g., Vega, D3, SVG) using a combination of direct extraction and LLM-assisted conversion. These representations are then compiled into raster images to ensure consistency across representations. Finally, we author chart descriptions and annotation tasks based on the chart content.

For professional charts, the process begins with collecting chart images from real-world sources and reconstructing their underlying data and structure. We implement a canonical version of each chart in D3.js, from which additional representations are derived. Similar to the library pipeline, we generate raster outputs and create corresponding chart descriptions and annotation tasks. This two-track pipeline ensures that AnnoBench contains both reproducible, structured examples and visually rich, real-world charts.

\section{Annotation Evaluation Examples}

\begin{figure*}
    \centering
    \includegraphics[width=\linewidth]{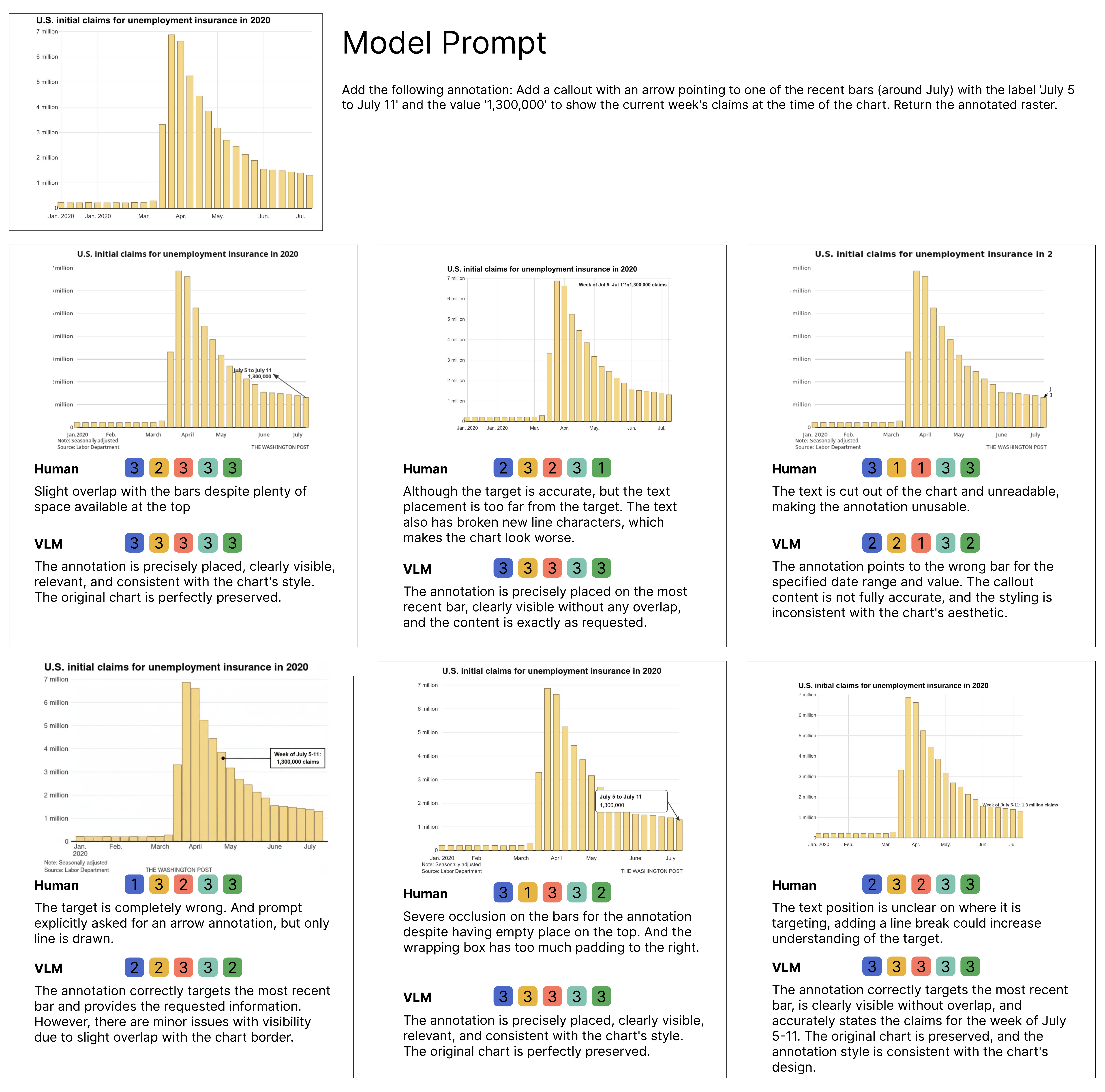}
    \caption{We provide a list of examples of an input chart, the annotation task prompt provided to the model, the model's output, and the evaluation on the output performed by a human and a VLM-Judge. Below the scorings, we provide human explanation and the VLMs explanation for the scoring that they did.}
    \label{fig:annotation-examples}
\end{figure*}

\autoref{fig:annotation-examples} presents representative examples of model outputs and their evaluation under our rubric. Each example includes the input chart, the annotation task prompt provided to the model, and the resulting annotated output. For each output, we report both human evaluation scores and VLM-as-a-judge scores across all rubric dimensions, along with brief explanations that justify the assigned scores.

These examples highlight common success cases and failure modes across models. In particular, they illustrate how errors may arise from incorrect target selection, inaccurate annotation content, poor spatial placement, or stylistic inconsistencies, even when other aspects of the annotation are correct. They also demonstrate differences between human and VLM-based evaluation, where visually plausible outputs may still fail to align with the underlying data or task requirements.

Together, these examples provide qualitative insight into how the evaluation framework captures annotation quality and how different models behave across a range of annotation scenarios.

\begin{figure*}
    \centering
    \includegraphics[width=\linewidth]{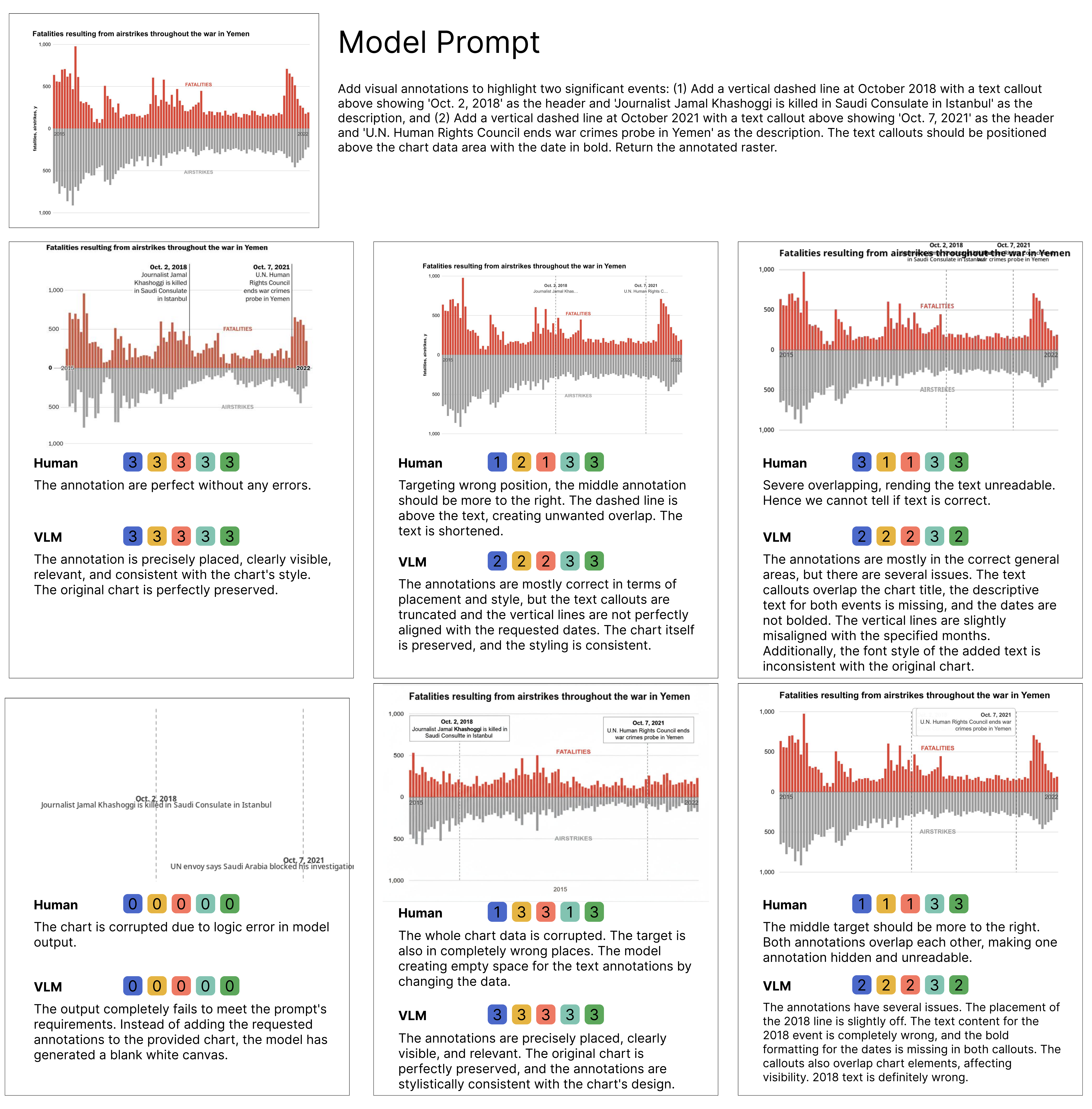}
    \caption{\autoref{fig:annotation-examples} continued.}
\end{figure*}
\begin{figure*}
    \centering
    \includegraphics[width=\linewidth]{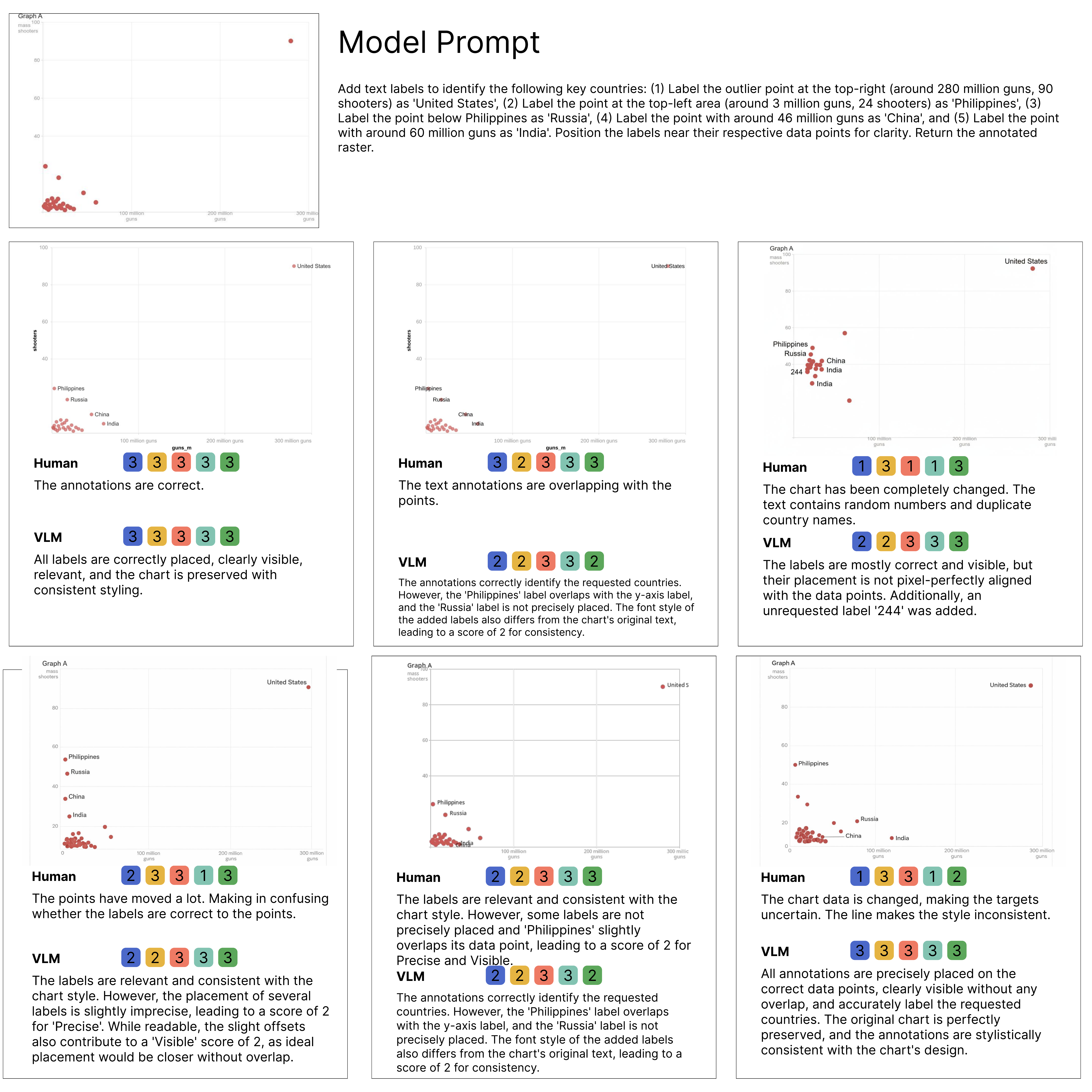}
    \caption{\autoref{fig:annotation-examples} continued.}
\end{figure*}
\begin{figure*}
    \centering
    \includegraphics[width=\linewidth]{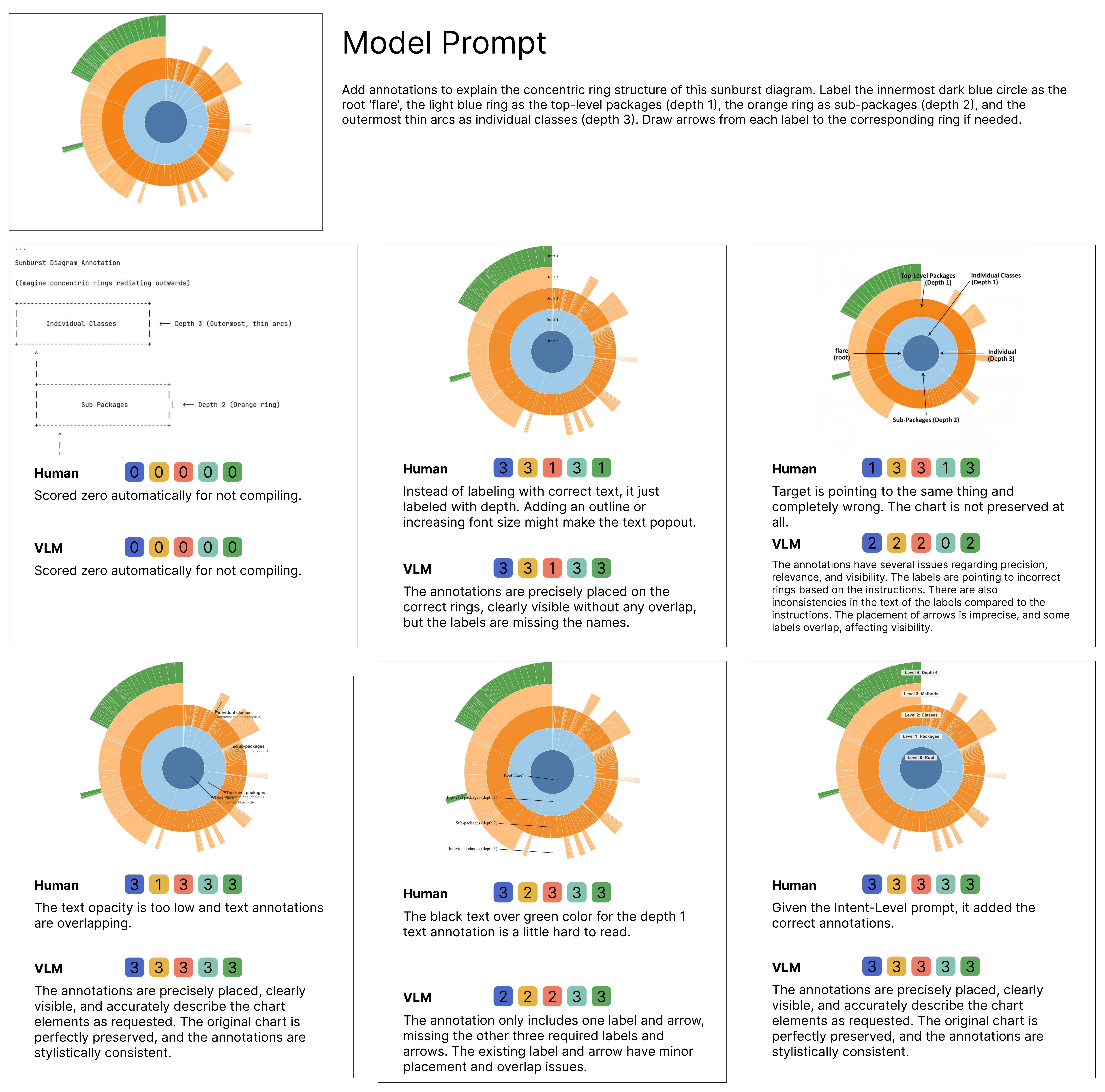}
    \caption{\autoref{fig:annotation-examples} continued.}
\end{figure*}